\documentclass[aps,twocolumn,superscriptaddress,letterpaper]{revtex4}
\usepackage{graphicx}
\usepackage{amssymb,amsmath}
\usepackage{float}

\usepackage{setspace}
\usepackage{parskip}
\usepackage{xcolor}
\definecolor{midnightblue}{cmyk}{1,1,0,0.1}
\definecolor{forestgreen}{cmyk}{0.76,0,0.26,0.5}

\usepackage[a4paper]{geometry}
\usepackage{epsfig}

\usepackage{hyperref}
\hypersetup{
    bookmarks=true,         % show bookmarks bar?
    unicode=false,          % non-Latin characters in Acrobat’s bookmarks
    pdftoolbar=true,        % show Acrobat’s toolbar?
    pdfmenubar=true,        % show Acrobat’s menu?
    pdffitwindow=false,     % window fit to page when opened
    pdfstartview={FitH},    % fits the width of the page to the window
    pdftitle={My title},    % title
    pdfauthor={Author},     % author
    pdfsubject={Subject},   % subject of the document
    pdfcreator={Creator},   % creator of the document
    pdfproducer={Producer}, % producer of the document
    pdfkeywords={keyword1} {key2} {key3}, % list of keywords
    pdfnewwindow=true,      % links in new window
    colorlinks=true,       % false: boxed links; true: colored links
    linkcolor=midnightblue,          % color of internal links (change box color with linkbordercolor)
    citecolor=magenta,        % color of links to bibliography
    filecolor=midnightblue,      % color of file links
    urlcolor=midnightblue,          % color of external links
}

\begin{document}

\title{Giant and tunable valley degeneracy splitting in MoTe$_2$}

\author{Jingshan Qi}
\altaffiliation{Equal contribution.}
\affiliation{School of Physics and Electronic Engineering, Jiangsu Normal University, Xuzhou 221116, P. R. China}

\author{Xiao Li}
\altaffiliation{Equal contribution.}
\affiliation{Department of Physics, University of Texas at Austin, Austin, TX 78712, USA}

\author{Qian Niu}
\email{niu@physics.utexas.edu}
\affiliation{Department of Physics, University of Texas at Austin, Austin, TX 78712, USA}
\affiliation{International Center for Quantum Materials, School of Physics, Peking University, Beijing 100871, P. R. China }
\affiliation{Collaborative Innovation Center of Quantum Matter, Beijing, P. R. China }

\author{Ji Feng}
\email{jfeng11@pku.edu.cn}
\affiliation{International Center for Quantum Materials, School of Physics, Peking University, Beijing 100871, P. R. China }
\affiliation{Collaborative Innovation Center of Quantum Matter, Beijing, P. R. China }

\begin{abstract}
Monolayer transition-metal dichalcogenides possess a pair of degenerate helical valleys in the band structure that exhibit fascinating optical valley polarization. Optical valley polarization, however, is limited by carrier lifetimes of these materials. Lifting the valley degeneracy is therefore an attractive route for achieving valley polarization. It is very challenging to achieve appreciable  valley degeneracy splitting with applied magnetic field. We propose a strategy to create giant splitting of the valley degeneracy by proximity-induced Zeeman effect. As a demonstration, our first principles calculations of monolayer MoTe$_2$ on a EuO substrate show that valley splitting over 300 meV can be generated. The proximity coupling also makes interband transition energies valley dependent, enabling valley selection by optical frequency tuning in addition to circular polarization. The valley splitting in the heterostructure is also continuously tunable by rotating substrate magnetization. The giant and tunable valley splitting adds a readily accessible dimension to the valley-spin physics with rich and interesting experimental consequences, and offers a practical avenue for exploring device paradigms based on the intrinsic degrees of freedom of electrons.
\end{abstract}

\maketitle
\textcolor{forestgreen}{\emph{\textsf{Introduction}.}}---
As promising valleytronics\cite{Xiao07,Akhmerov07,Rycerz07,Yao08,Li13,Li13d,Li14a} materials, monolayer transition-metal dichalcogenides, MX$_2$ (M = Mo, W and X=S, Se and Te), have attracted significant research interest in recent years~\cite{Cao12,Xiao12,Zeng12,Mak12}. The fascinating optoelectronic properties arise from a pair of degenerate but inequivalent valleys in the vicinities of the vertices of the hexagonal Brillouin zone (BZ). The principal challenge in exploiting the valley degree of freedom in electronic and optoelectronic applications is the ability to generate and control the valley polarization. Dynamical polarization of monolayer MX$_2$ has been achieved by circularly polarized optical pumping,~\cite{Cao12,Xiao12,Zeng12,Mak12} based on valley-selective circular dichroism; that is, the interband transitions at two valleys exclusively absorb the left- and right-polarized photon, respectively~\cite{Yao08,Cao12,Xiao12}. The subsequent Berry curvature-induced valley Hall effect of MoS$_2$ has also been observed in a recent experiment.~\cite{Mak14} These findings offer a new paradigm for novel optoelectronic devices.

Dynamical valley polarization based on optical pumping relies upon transient non-equilibrium photo-carrier distribution, and is ultimately limited by carrier lifetime. Although the fact that photo-carriers are rather short-lived for MX$_2$ (see \cite{Wang13,Lagarde14} and references therein) is more an issue with material quality than a fundamental limit, it does present a challenge to utilizing valley polarization experimentally, especially in beyond-proof-of-concept devices. An alternative route to valley polarization is lifting the valley degeneracy by breaking time-reversal symmetry. Valley degeneracy splitting will make possible equilibrium valley polarization via conventional doping, although schemes for creating significant valley splitting has yet to be developed. Valley splitting in MX$_2$ has been assessed in a few recent experiments, which showed that only small valley splitting, 0.1--0.2 meV/tesla, can be generated by an external magnetic field.~\cite{Li14c,MacNeill15,Aivazian15,Srivastava15} Here, we propose a strategy to create giant valley splitting by proximity-induced magnetic interactions. We use first principles calculation to demonstrate that in molybdenum ditelluride (MoTe$_2$) on a europium  oxide (EuO) substrate valley splitting larger than 300 meV can be generated by an induced Zeeman field. Additional Rashba field induced by the substrate and Zeeman field further splits the degeneracy in the interband optical transition between like spins. Based on the lifting of valley and spin degeneracies, we discuss new avenues for exploring novel spin and valley physics by equilibrium doping or dynamical optical excitation. We also demonstrate that the valley splitting resulting from the proposed strategy is highly tunable.

\textcolor{forestgreen}{\emph{\textsf{MoTe$_2$/EuO heterostructure}.}}---
The structure of monolayer MX$_2$ is shown in Fig. 1(a), which mimics a two-dimensional honeycomb lattice. The corresponding BZ and high symmetry points are depicted in Fig. 1(b). The inequivalent $K_\pm$ are the vertices of the hexagonal BZ, where the direct band gaps are located. Typical low-energy band structure of free-standing monolayer MX$_2$ is shown in Fig. 1(c). In the absence of inversion symmetry, the bands are not spin degenerate and the electronic states with up/down spins are represented by blue/red lines with up arrows ($\uparrow$)/down arrows ($\downarrow$). We now introduce a first energy scale pertinent to the important valley-spin physics. The spin splitting within a valley (see Fig. 1(c)), 
$\Delta_{\mathrm{spin}}^{v/c,\tau}\equiv E_{\uparrow}^{v/c,\tau} - E_{\downarrow}^{v/c,\tau}$, 
arising from spin-orbit coupling (SOC). Here, $v$ and $c$ refer respectively to the valence and conduction bands. Valleys, $K_\pm$, are addressed by the index $\tau=\pm 1$. $E$ is the corresponding energy level. When the system is invariant under time reversal, the spin splitting is opposite for two valleys. Because different  $d$-orbitals dominate the valence and conduction bands,~\cite{Cao12}  the magnitude of the spin splitting of the valence bands ($\sim$ 150-500 meV) is typically larger than that of the conduction bands ($\sim$ 3-60meV).~\cite{Liu13} The relative signs of $\Delta_{\mathrm{spin}}^{c/v,\tau}$ for the conduction and valence bands depends only on the type of the metal atom:  the signs are opposite for MoX$_2$ (as is the case in Fig. 1, where $\Delta_{\mathrm{spin}}^{v,+}$ is set to be positive), while the signs are the same for WX$_2$.

\begin{figure}[h!]
\centering
\includegraphics[scale=1]{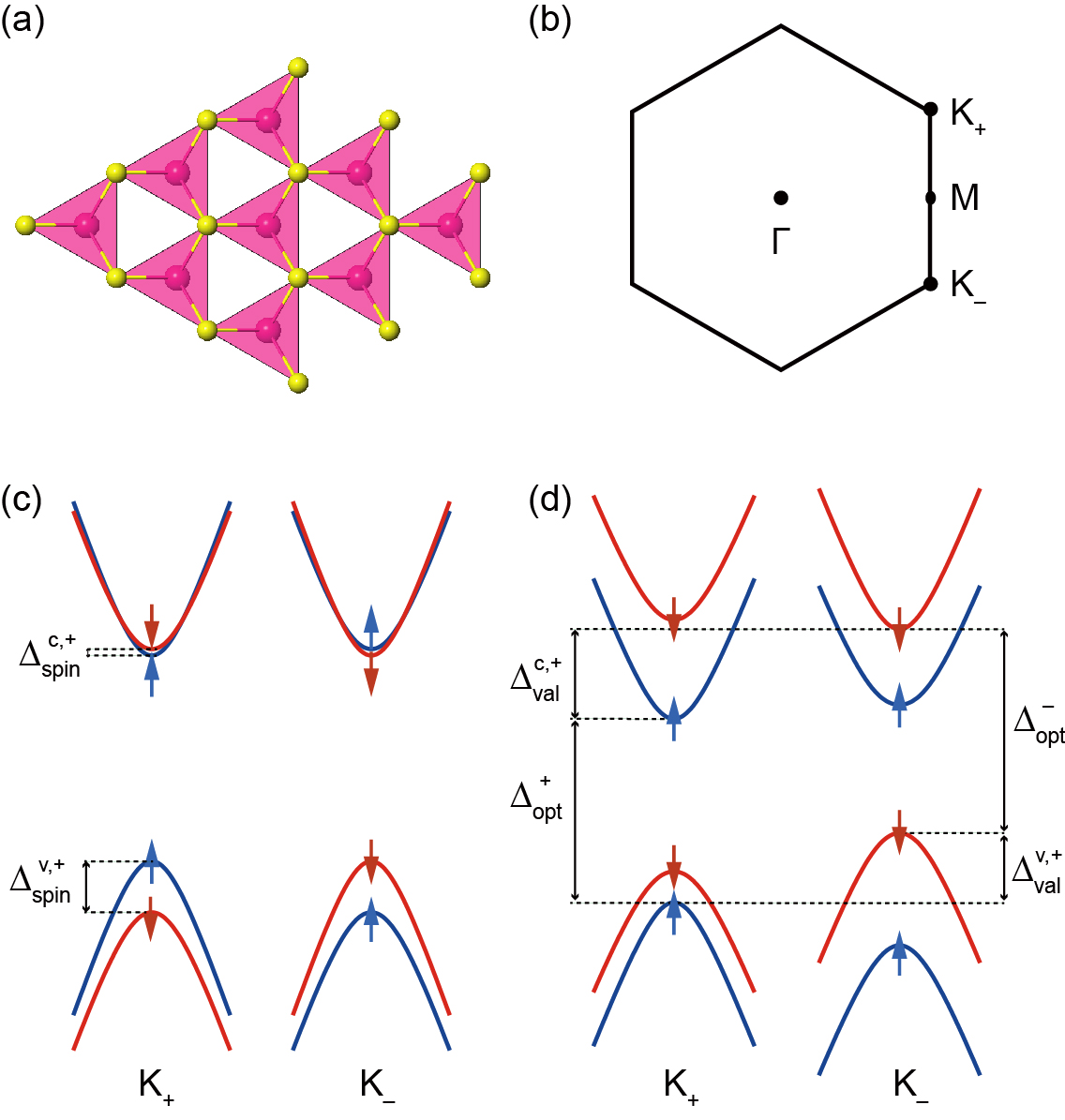}
\caption{Atomic and schematic band structures of monolayer MX$_2$. (a) Top view of a monolayer MX$_2$, which mimics a bipartite honeycomb lattice. The two lattice sites of the honeycomb lattice are occupied by M (pink spheres) and a pair of X (yellow spheres). Each M is coordinated to six X atoms forming a trigonal prism (pink triangles).  (b) The Brillouin zone and high symmetry points. (c) and (d) Schematic band structure at the two valleys of monolayer MX$_2$ without/with a Zeeman field, respectively. Blue bands with blue up arrows correspond to the up spin states, and red bands with red down arrows the down spin states.}
\label{fig:1}
\end{figure}

The valley degeneracy can be quantified by another energy scale, $\Delta_{\mathrm{val}}^{c/v,\tau}  \equiv E_{\uparrow}^{c/v,\tau}-E_{\downarrow}^{c/v,-\tau}$, which is zero in the presence of time-reversal symmetry. This important valley degeneracy is fundamental to the notion of valley degree of freedom and valleytronics based on dynamical valley polarization with chiral optical pumping induced non-equilibrium carrier distribution~\cite{Xiao07, Yao08, Cao12, Xiao12}. It is on the other hand also highly desirable to break the valley degeneracy to exploit the valley and spin degrees of freedom, with equilibrium doping or dynamical optical excitation, or the combination of both.~\cite{Li13} The application of external magnetic field breaks time-reversal symmetry and lifts the valley degeneracy, whereby $\Delta_{\mathrm{val}}^{c/v,\tau}\neq 0$, as exemplified in Fig. 1(d). Systematic measurements revealed that the Zeeman splitting of valley degeneracy only amounts to $\sim 0.1$ meV/T.~\cite{Li14c} Evidently, it is of importance to develop alternative strategies to achieve large valley splitting.

A promising approach is to employ the magnetic interaction to lift the valley degeneracy in heterostructure composed of monolayer MX$_2$ and an insulating ferromagnetic substrate. In the followings, by first-principles density functional theory (DFT) calculations, we take MoTe$_2$/EuO(111) as an example to demonstrate a giant valley splitting via proximity-induced magnetic interaction. In our first-principles Density functional theory calculations, the Perdew-Burke-Ernzerhof exchange-correlation functional and the projector-augmented wave potentials are used, as implemented in the VASP code.~\cite{Kresse96}  An energy cutoff of 400 eV for the plane-wave basis is adopted. We use a $15\times15\times15$ Monkhorst-Pack $\bm{k}$-mesh for bulk EuO, while a $15\times15\times1$ $\bm{k}$-mesh for the pristine MoTe$_2$ monolayer and MoTe$_2$/EuO. For MoTe$_2$ monolayer and MoTe$_2$/EuO, a vacuum slab of 20 $\mathrm{\AA}$ is inserted to minimize the interaction between the slab model and its periodic images. Atomic coordinates are optimized with a convergence threshold of 0.01 eV/$\mathrm{\AA}$ on the interatomic forces. We use GGA+$U$ calculation to account for on-site Coulomb interactions.~\cite{Dudarev98} The Coulomb and exchange parameters, $U$ and $J$, are set to 8.3 eV and 0.77 eV for Eu $4f$ orbitals, respectively, while  $U=4.6$ eV and  $J=1.2$ eV for O $2p$ orbitals.~\cite{Yang13,Ingle08} These values give a good agreement with the experimental results of EuO.~\cite{Steeneken02,Moodera07} The Kohn-Sham Bloch wavefunctions on the discretized $\bm{k}$-mesh are extracted to calculation the optical interband transition matrix elements~\cite{Cao12, Li13} and Berry curvature~\cite{Fukui05}.

We choose MoTe$_2$/EuO heterostructure as an example,  based on two considerations. First, EuO is a ferromagnetic semiconductor with a large band gap of more than 1 eV, and offers exchange interaction with $\sim 7\;\mu_{\mathrm{B}}$ spin moment on each Eu ions.~\cite{McGuire64,Schmehl07} Second, EuO(111) single-crystal film has very recently been grown on Ir(111)~\cite{Schumacher14}, and the lattice mismatch between MoTe$_2$ and EuO (111) substrate is only 2.7\%, a reasonable value for a commensurate heterostructure composed of slightly strained MoTe$_2$ monolayer on EuO(111). The calculated lattice constant of cubic EuO is 5.188 $\mathrm{\AA}$ and in-plane lattice constant of EuO (111) substrate is therefore set to 3.66 $\mathrm{\AA}$, which is very close to  the calculated lattice constant of free-standing MoTe$_2$ monolayer,  3.56 $\mathrm{\AA}$. 

We therefore construct the MoTe$_2$/EuO heterostructure with a MoTe$_2$ monolayer placed on the Eu-terminated surface of EuO (111) substrate composed of 12 Eu/O atomic layers, as shown in Fig. 2(a). The oxygen-terminated surface of EuO substrate is saturated by hydrogen, to model a semi-infinite single-crystalline EuO or EuO film grown on another substrate\cite{Schumacher14}. Structural relaxation reveals a few stable configurations for this MoTe$_2$/EuO heterostructure, corresponding to relative shifts of the two materials along the (111) plane of EuO. As detailed in the Supplementary Information (SI), we find that one of these configurations to be substantially more stable than the others, which will be focused upon in subsequent discussions. In this configuration, Mo atoms are located directly on top of Eu ions, favouring proximity-induced magnetic effects (Figs. 2 (a) and (b)). 

% Moreover, the calculated band gap of the unstrained MoTe$_2$ monolayer is 0.94 eV, which is smaller than EuO's gap 1.1 eV. 

\begin{figure}[h!]
\centering
\includegraphics[scale=0.95]{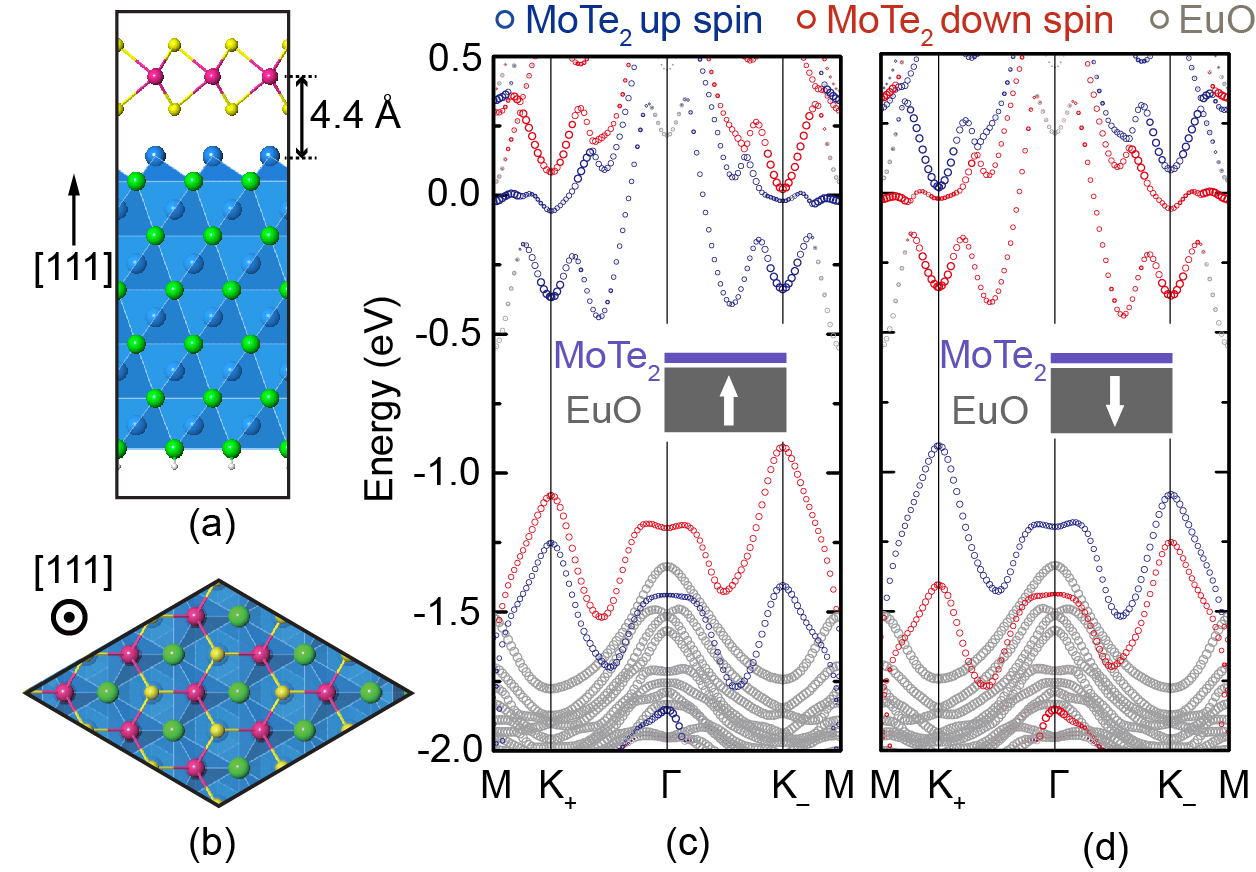}
\caption{Atomic and band structures of MoTe$_2$/EuO system. (a) Side view and (b) top view of MoTe$_2$ monolayer on EuO (111) substrate. (c) and (d) Band structures of MoTe$_2$/EuO with EuO magnetized upward and downward, respectively. We use fat band representation to indicate the projected weights on MoTe$_2$ (blue and red) and EuO (gray). The MoTe$_2$ projections are colored according to the direction of spin magnetic moment near K$_\pm$: blue for up spin and red for down spin. The energy scale is zeroed to the Fermi level.}
\label{fig:2}
\end{figure}

Fig. 2(c) and (d) show the calculated band structures of MoTe$_2$/EuO heterojunction, with EuO magnetized in the upward and downward directions, respectively (see insets). The fat-band representation corresponds to the projected weights of each band onto MoTe$_2$ (blue and red bands), and EuO (gray bands). We see that a few bands arise primarily from MoTe$_2$ in the energy range between -1.5 and 0.5 eV, where EuO substrate has only minor contribution. Viewing the MoTe$_2$ bands only, there is a well-defined global gap in the energy range between -0.9 and -0.4 eV. The direct gaps at $K_\pm$ ($\sim 0.6-0.7$ eV) indeed corresponds to valleys of native free-standing monolayer MoTe$_2$. This is also supported by the computed optical selectivity and Berry curvature, to be presented shortly.

The identification of MoTe$_2$ bands largely free of hybridization with the substrate leads immediately to the key observation that the valley degeneracy of MoTe$_2$ bands is substantially lifted. The MoTe$_2$ bands near the gap can be classified as spin up and down, respectively colored in blue and red in Fig. 2 (c) and (d), as the projected spin magnetic moments near K$_\pm$ are dominantly out-of-plane, along the [111] direction of EuO (see Fig. S2 of SI). Therefore, the valley splitting can be quantified by the magnitude of $\Delta_{\mathrm{val}}^{c/v,\pm}$, which are in the range from 321 to 419 meV (see Table \ref{table:1}). Moreover, the smallest energies for the band edge vertical optical transition without spin flip become unequal, with $\Delta_{\mathrm{opt}}^+ = E_{\uparrow}^{c,+} - E_{\uparrow}^{v,+}= 886$ meV and $\Delta_{\mathrm{opt}}^- = E_{\downarrow}^{c,-} - E_{\downarrow}^{v,-}=930$ meV (Fig. 2 (c)), and the difference  $\Delta_{\mathrm{opt}}^- - \Delta_{\mathrm{opt}}^+$ reaches a substantial value 44 meV, equivalent to the splitting by a 440-tesla magnetic field.

\textcolor{forestgreen}{\emph{\textsf{Proximity-induced magnetic interactions}.}}---
It is evident that the substrate induces substantial magnetic interaction by proximity effect, and leads to giant valley splitting in MoTe$_2$. In order to gain further insight into the proximity-induced magnetic interactions, a low-energy effective Hamiltonian (LEH) is constructed to fit the band structure of MoTe$_2$/EuO in the vicinity of $K_\pm$, showing that Zeeman and Rashba fields are the key interactions induced by the proximity effect. The LEH is composed of four parts, $H=H_0+H_{\mathrm{soc}}+H_{\mathrm{ex}}+H_{\mathrm{R}}$. The four terms correspond respectively to the orbital interactions, SOC-induced spin-splitting, proximity-induced exchange and Rashba interactions, given as
\begin{subequations}
\begin{align}
H_0 &= v_{\mathrm{F}}(\tau\sigma_xp_x+\sigma_yp_y)+\frac{m}{2}\sigma_z;
\label{eq:1a}
\\
H_{\mathrm{soc}} &= \tau s_z(\lambda_c \sigma_+ + \lambda_v \sigma_-);
\label{eq:1b}
\\
H_{\mathrm{ex}} &= -s_z(B_c\sigma_+ + B_v\sigma_-);
\label{eq:1c}
\\
H_{\mathrm{R}} & = \lambda_{\mathrm{R}} (\tau s_y \sigma_x-s_x \sigma_y).
\label{eq:1d}
\end{align}
\end{subequations}
Here, the Pauli matrices $s_\alpha$ and $\sigma_\alpha$ ($\alpha=0,x,y,z$) refer to real spin and orbital pseudo spin, respectively, and $\sigma_\pm\equiv \frac{1}{2}(\sigma_0\pm\sigma_z)$.  $\bm{p}$ is the electronic momentum and $v_{\mathrm{F}}$ is the Fermi velocity. The spin splitting of the conduction and valence bands due to intrinsic SOC is determined by parameters $\lambda_c$ and $\lambda_v$, respectively. The effective mass, $m$, corresponds to the band gap in the absence of SOC, which results mainly from the crystal-field splitting between $d_{z^2}$ and $\{d_{xy}, d_{x^2-y^2}\}$ of Mo. $B_{c}$ and $B_v$ are effective Zeeman fields experienced by the conduction and valence bands of MoTe$_2$, arising from the coupling with a magnetic substrate. Note that the low-energy band structure of free-standing monolayer MX$_2$, as shown schematically in Fig. \ref{fig:1}(c), can be described by $H_0+H_{\mathrm{soc}}$ (Eqs. (\ref{eq:1a}) and (\ref{eq:1b}) together).~\cite{Xiao12}

\begin{table}
\caption{Important energy scales of valley and spin, for MoTe$_2$ monolayer and MoTe$_2$/EuO heterostructure.}

\begin{center}
\begin{tabular}{l|rrrr}

\hline
Valley splitting (meV)& 
$\Delta_{\mathrm{val}}^{v,+}$ &
$\Delta_{\mathrm{val}}^{c,+}$ &
$\Delta_{\mathrm{val}}^{v,-}$ &
$\Delta_{\mathrm{val}}^{c,-}$ 
 \\
\hline

MoTe$_2$$^\dagger$  (DFT \& LEH)   & 
0 &
0& 
0&
0
 \\
MoTe$_2$/EuO (DFT)& 
-342&
-386&
-321&
-419 \\
MoTe$_2$/EuO (LEH)& 
-319&
-412&
-340&
-391 \\
\hline
\hline
Spin splitting (meV)& 
$\Delta_{\mathrm{spin}}^{v,+}$ &
$\Delta_{\mathrm{spin}}^{c,+}$ &
$\Delta_{\mathrm{spin}}^{v,-}$ &
$\Delta_{\mathrm{spin}}^{c,-}$ 
 \\
\hline
MoTe$_2$$^\dagger$ (DFT \& LEH)     & 
214 &
-27& 
-214&
27
 \\
MoTe$_2$/EuO (DFT) & 
-168&
-449&
-496&
-356 \\
MoTe$_2$/EuO (LEH) & 
-142&
-455&
-517&
-348 \\
\hline
\end{tabular}

$^\dagger$ For free-standing MoTe$_2$ monolayer, DFT calculation and LEH give the same results.
\end{center}
\label{table:1}
\end{table}

$H_{\mathrm{ex}}$ represents the Zeeman field induced by the substrate, and produces a band structure schematically shown in Fig. \ref{fig:1}(d). The valley degeneracy is clearly broken by the Zeeman field, with $\Delta_{\mathrm{val}}^{c/v,\tau}=-2 B_{c/v}$, which is independent of the valley index $\tau$. This, however, is inconsistent with the DFT results (see Table \ref{table:1}), for which $\Delta_{\mathrm{val}}^{c/v,\tau}$ is dependent on the valley index. To account for the $\tau$-dependence of $\Delta_{\mathrm{val}}^{c/v,\tau}$ and considering the surface electric field in the (111)-direction, we include the Rashba field embodied in $H_{\mathrm{R}}$.~\cite{Ochoa13} The Rashba effect further hybridizes the valence and conduction bands, and mixes the spin components, with which the spin is no longer a good quantum number. However, we can still designate the low-energy bands of MoTe$_2$ monolayer with up/down spin, since the out-of-plane spin components dominates in the valley region (Fig. S2 of SI). With the Rashba term is added,  $\Delta_{\mathrm{val}}^{c/v,\tau}$  become valley-dependent, in agreement with first principles band structure. Based on this model (Eqs. (\ref{eq:1a})-(\ref{eq:1d}) together), we can deduce that $\Delta_{\mathrm{val}}^{c,\tau}-\Delta_{\mathrm{val}}^{v,-\tau}=2(B_v-B_c)$. From the values in Table \ref{table:1}, the effective Zeeman fields for the conduction and valence bands are seen to differ by as much as 30-40 meV, justifying the use of two effective Zeeman fields, $B_{c/v}$, which reflect different effective Land\'{e} $g$-factors for the Bloch states.

Matching the model with the first principles band structure of MoTe$_2$/EuO leads to a semiquantitative clarification of the role of proximity-induced magnetic interactions. First, we obtain the parameters, $v_{\mathrm{F}}=3.75 \times 10^5$ m/s, $\lambda_c=-13$ meV, $\lambda_v=107$ meV and $m=861$ meV, in the low-energy model of free-standing MoTe$_2$ monolayer  (i.e. $H_0+H_{\mathrm{soc}}$), by fitting the corresponding DFT band structure. With EuO substrate is added, we keep $\lambda_{c/v}$ unchanged and all other parameters in Eq. 1 are refined by a least-square fit to MoTe$_2$-projected bands near $K_\pm$ in Fig. 2(c) (detailed in SI). The Fermi velocity is largely unaffected by the substrate, with $v_{\mathrm{F}} = 3.95\times10^5$ m/s. The effective mass is significantly renormalized to a value 984 meV. The effective Zeeman fields are gigantic: $B_c = 206$ meV, $B_v=170$ meV, which translates to a magnetic field over 2937 tesla. For a comparison, the magnetic field produced by the magnetic moments in Eu ions 4.4 $\mathrm{\AA}$ above the EuO(111) surface is only $\sim$0.007 tesla, which is computed from the direct summation of the magnetic field produced by magnetic dipoles of Eu ions (7 $\mu_B$) in a semi-infinite EuO crystal. The Rashba parameter is determined to be $\lambda_{\mathrm{R}}=72$ meV. The relevant energy scales of the LEH model with as-determined parameters above are summarized in Table \ref{table:1}, and it is seen that they are in decent agreement with the DFT results. It can be concluded that the proximity-induced spin interaction is dominated by gigantic Zeeman effect, with significant Rashba interaction.

\textcolor{forestgreen}{\emph{\textsf{Discussions}.}}---
The valley splitting by proximity-induced magnetic interaction is attractive, as it creates giant difference in various energy scales between the valleys. In this section we discuss how the giant valley splitting will facilitate the access and manipulation of valley and spin degrees of freedom in a MoTe$_2$/EuO heterostructure. In order to access the valleys of MoTe$_2$, we will need to tune the chemical potential into the MoTe$_2$ gap (between  $\sim -0.9$ and $-0.4$ eV), which may be achieved with electrical gating or chemical doping. In the ensuing discussions, we will assume that we begin with a system of which the chemical potential is already tuned into the MoTe$_2$ gap. In what follows, we will discuss essential experimental consequences of the proposed valley degeneracy splitting, in terms of transport and optical properties, and tunability of valley splitting.

\begin{figure}[h!]
\centering
\includegraphics[scale=1]{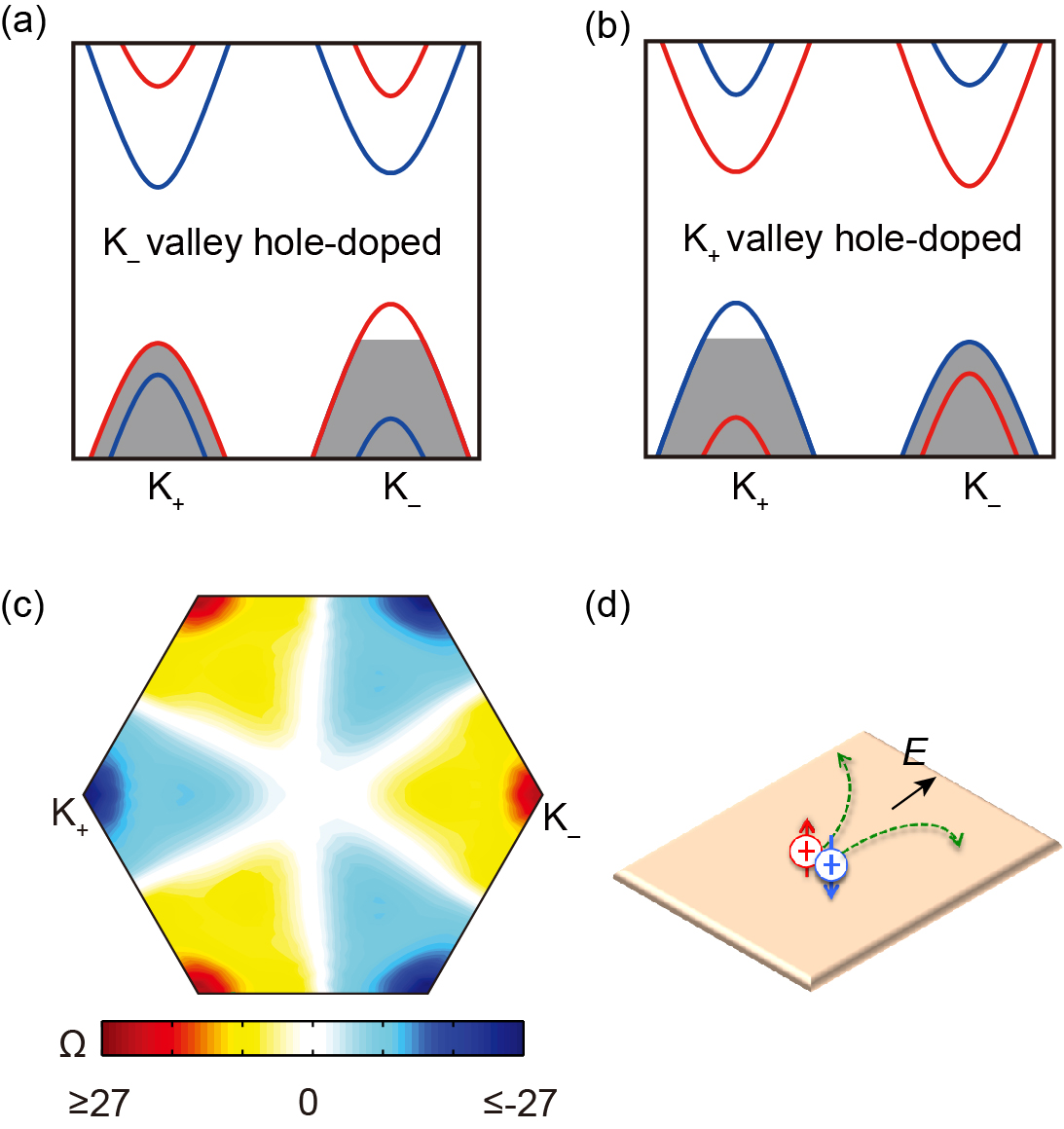}
\caption{Anomalous valley transport of hole-doped MoTe$_2$/EuO heterostructure. (a) and (b) show the schematic hole-doped MoTe$_2$-projected band structure of the heterostructure when the substrate EuO is magnetized upward and downward, respectively (\emph{c.f.} Figs. \ref{fig:2} (c) and (d)). (c) Computed non-Abelian Berry curvature~\cite{Fukui05}, $\Omega$ (units $\mathrm{\AA}^2$), of the valence bands occupied up to the MoTe$_2$ gap of the heterostructure when EuO substrate is magnetized upward (band structure in Fig. \ref{fig:2}(c)). The Berry curvature is momentum-resolved and shown in the entire Brillouin zone. (d)  Schematic depiction of  anomalous Hall effects.  The positive sign in circles denotes the hole. Holes with red up and blue down arrows denote up-spin one from $K_-$ valley in (a) and down-spin one from $K_+$ valley in (b), respectively. $\bm{E}$ is the applied in-plane electric field.}
\label{fig:3}
\end{figure}

Owing to valley splitting, we can selectively dope only one valley with carriers at equilibrium. Based on the band structure in Fig. \ref{fig:2}(c), it is seen that hole doping will give simple access to the valley states of MoTe$_2$ monolayer, whereas electron doping will be interfered by the substrate's bands. We will therefore focus on the case with hole doping. If we dope the heterostructure with holes at $K_+$ valley when the substrate is magnetized upward (Fig. \ref{fig:3}(a)), the spin holes may produce a transversal current under a  longitudinal in-plane electric field (Fig. \ref{fig:3}(d)).~\cite{Xiao10} The anomalous transport arises from the \emph{intrinsic} Hall conductivity without magnetic field
\begin{equation}
\sigma_{xy} = \frac{e^2}{\hbar} \int \frac{dk^2}{(2\pi)^2} \Omega(\bm{k})n(\bm{k}),
\end{equation}
 where $\Omega(\bm{k})$ is the Berry curvature of Bloch electron~\cite{Xiao10}, and $n(\bm{k})$ the occupation of $\bm{k}$-space (from doping or otherwise), $e$ the elementary charge and $\hbar$ the reduced Planck's constant. This anomalous Hall conductivity is an intrinsic property of the Bloch band, and is the key to the detection of valleys by electric measurement~\cite{Xiao07,Yao08}. It therefore is desirable to have sizable  $\Omega(\bm{k})$ especially in the vicinity of valleys of of MoTe$_2$ monolayer. 
 
 As shown in Fig. \ref{fig:3}(c), the calculated Berry curvature is sharply peaked in the valley region, with opposite signs for $K_\pm$. Clearly, the valley identity remains intact apart from the giant valley splitting and shall lead a transverse current under an applied electric field. The flux of the spin holes carries three observable physical quantities, namely, charge, spin moment and valley-dependent orbital magnetic moment, which can be referred to respectively as anomalous charge, spin and valley Hall effects. Due to the dominance of the Zeeman effect as discussed above, it is expected that when the magnetization of the substrate is flipped, the magnetically induced valley polarization will correspondingly be flipped. This confirmed by the DFT calculation, as shown by the band structure in Fig. \ref{fig:2}(d). Experimentally, this corresponds to switching the magnetization of EuO substrate with a relatively small magnetic field, as EuO has very weak magnetic anisotropy~\cite{Miyata67}. With the reversed magnetization, the valley polarization as well as the associated anomalous charge/spin/valley Hall effects are reversed (Figs. \ref{fig:3}(b) and (d)). 

\begin{figure}[h!]
\centering
\includegraphics[scale=1]{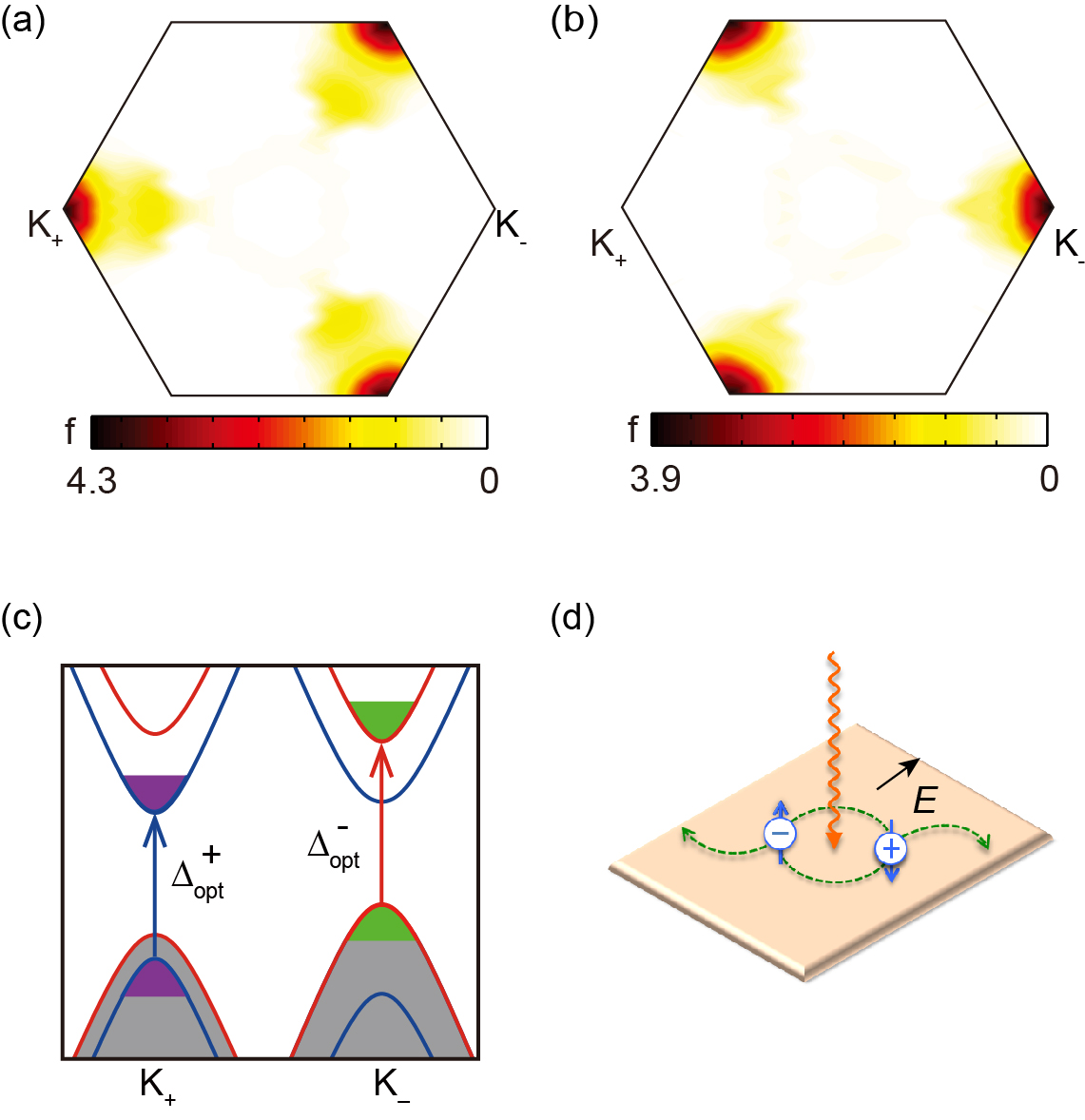}
\caption{Dynamical valley polarization in MoTe$_2$/EuO heterostructure. (a) and (b) show the $\bm{k}$-resolved oscillator strength between the up-spin valence and conduction bands, $f(\bm{k})$, throughout the Brillouin zone for left- and right-polarized lights, respectively. The oscillator strength between the down-spin valence and conduction bands is similar and not shown. (c) Schematic depiction of optical pumping in MoTe$_2$/EuO heterostructure with EuO magnetized toward MoTe$_2$ monolayer. The lowest-energy interband transition of the two valleys with energies of $\Delta_{\mathrm{opt}}^+$ and $\Delta_{\mathrm{opt}}^-$ are indicated. (d) Optically induced Hall effects.  The positive and negative signs in circles denote the photoexcited hole and electron, respectively. Up and down arrow denotes up spin and down spin respectively. $\bm{E}$ is the applied in-plane electric field.}
\label{fig:4}
\end{figure}

The valley identity is also associated with valley-contrasting circular dichroism in optical absorption in addition to the contrasting anomalous transport, which is a very important property of MX$_2$ monolayer. The circular dichroism is readily characterized by the optical oscillator strength under the circularly polarized optical field, $f_\pm(\bm{k}) = |\mathcal{P}_\pm(\bm{k})|^2/m_e(E_{f,\bm{k}}-E_{i,\bm{k}})$, where $\mathcal{P}_\pm=\mathcal{P}_x\pm i \mathcal{P}_y$ with $\mathcal{P}_\alpha=\langle \psi_f|p_\alpha|\psi_i\rangle$ ($\alpha=x,y$) being the interband matrix element, $m_e$ is the electron mass, $E_i$ and $E_f$ are the energies of initial and final Bloch states, and $\psi_i$, $\psi_j$ are the corresponding wavefunctions.   As shown in Fig. \ref{fig:4}(a) and (b), despite the giant splitting of valley degeneracy the optical absorption of the MoTe$_2$/EuO still preserves perfect circular dichroism  in the vicinity of $K_\pm$ valleys. The optical oscillator strength for left-polarized light is sharply peaked near $K_+$, but is vanishingly small near $K_-$, and vice versa for right-polarized light, which will allow easy chiral optical pumping induced  valley polarization in EuO-supported MoTe$_2$ monolayer, similar to free-standing MoTe$_2$ monolayer.

 The proximity-induced Zeeman and Rashba effects together lifts the optical frequency degeneracy, as discussed above, which adds an interesting variable to the dynamical optical valley polarization. For free-standing MoTe$_2$ monolayer with degenerate valleys, non-polarized and linearly polarized light excite carriers equally from both valleys, and \emph{cannot} lead to valley polarization. In contrast, dynamical valley polarization can be achieved in EuO-supported MoTe$_2$ monolayer by selection rule based on energy conservation, with a photon energy $\hbar\omega$ satisfying the relation $\Delta_{\mathrm{opt}}^+\le\hbar\omega<\Delta_{\mathrm{opt}}^-$ for the EuO substrate  magnetized upward, regardless of light polarization. Due to the giant splitting of valley degeneracy, $\Delta_{\mathrm{opt}}^- - \Delta_{\mathrm{opt}}^+ = 44$ meV, the energy selection works over a wide spectral range. When light within the energy range above illuminates the sample, only electron-hole pairs from the $K_+$ valley are generated (Fig.  \ref{fig:4} (c)). The photoexcited electrons and holes may recombine, leading to  only left-polarized luminescence.  Under an in-plane electric field, the electrons and holes acquire opposite transversal velocities and move to opposite boundaries of the sample (Fig.  \ref{fig:4} (d)), due to the opposite signs of the Berry curvatures in the conduction and valence bands. This also leads to a net charge/spin/valley Hall current. With the reversed magnetization of the substrate, the carriers will be excited at $K_-$ valley  and the corresponding charge/spin/valley Hall effects are reversed.

The last and very important remark on the proposed valley splitting is concerning its tunability. As noted before, the magnetic anisotropy of EuO is weak~\cite{Miyata67}, and can be easily rotated with a relatively small external magnetic field. As this is a readily accessible experimental knob, it is useful to discuss how the valley splitting depends on the direction of substrate magnetization. As discussed above, the proximity-induced interaction is dominated by the Zeeman field, and we therefore ignore the change to the Rashba field as the magnetization rotates in a first approximation. A straightforward generalization of Eq. (\ref{eq:1c}) can be introduced to describe the field-tuning of substrate magnetization,
\begin{equation}
H_{\mathrm{ex}} = 
-\bm{s}\cdot\hat{\bm{n}} (B_c\sigma_+ + B_v\sigma_-),
\end{equation}
where $\hat{\bm{n}}$ is a unit vector denoting the direction of the proximity-induced Zeeman fields, which enters as $\bm{B}_{c/v}= \hat{\bm{n}}B_{c/v}$. One notes that $\bm{B}_{c/v}$ are the induced effective Zeeman field, rather than the applied magnetic field for rotating the magnetization of the substrate. The latter field is supposed to be small so as not to interfere with subsequent transport measurements. 

\begin{figure}[h!]
\centering
\includegraphics[width=75mm]{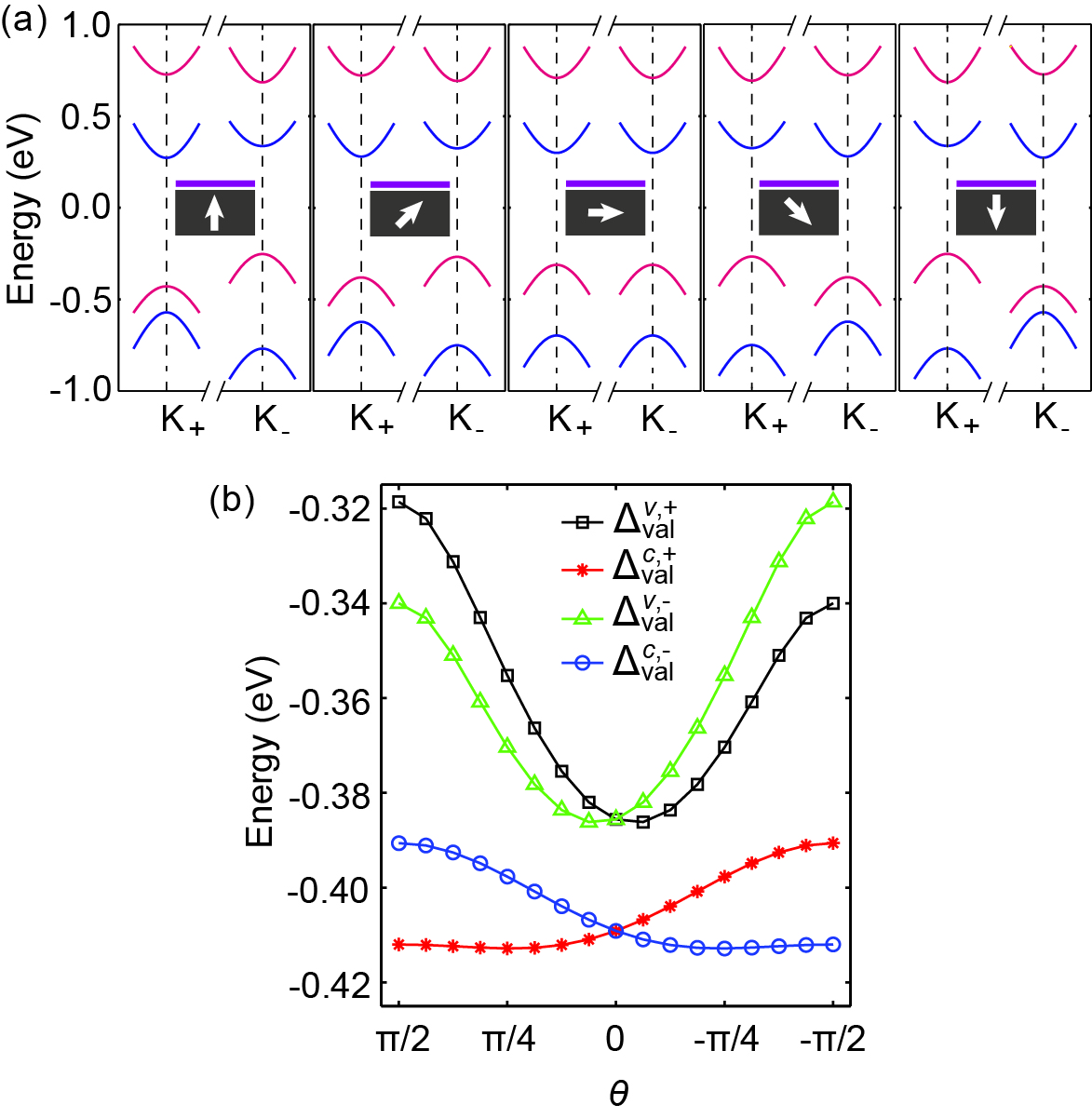}
\caption{Highly tunable valley splitting in MoTe$_2$/EuO heterostructure. (a) Band structure in the vicinity of the valleys (momentum range $\pi/5a$ around $K_{\pm}$, where $a$ is the lattice constant of MoTe$_2$) from the LEH model, for $\theta=\pi/2,\pi/4,0,-\pi/4,-\pi/2$. The bands are colored red and blue when the spins are antiparallel and parallel with the substrate magnetization (see insets), respectively. (b) Valley splittings as functions of the substrate magnetization direction.}
\label{fig:5}
\end{figure}

Taking $\hat{\bm{n}} = (\cos\theta,0,\sin\theta)$ and assuming all parameters of the model to remain unchanged, the corresponding evolution of MoTe$_2$ valley band structure is shown in Fig. \ref{fig:5}(a) to demonstrate the effects of field tuning of substrate magnetization. It is evident that as the magnetization of the substrate is rotated, the valley splitting can be continuously tuned. Compared to conduction bands, valence bands are more sensitive to the change of the magnetization direction. In Fig. \ref{fig:5}(b), the valley splittings (see caption) for conduction and valence bands are plotted as functions of $\theta$, as the magnetization rotates from upward, to horizontal, and finally to downward directions. As the magnetic field turns from perpendicular to horizontal direction, the change in valley splitting for valence bands is $\sim 46-67$ meV, whereas the counterpart for conduction bands is $\sim 3-19$ meV. The large change of valence valley splitting is clearly advantageous, as the valence bands in the valleys are unobstructed by substrate bands or the bands in other regions of the $\bm{k}$-space (see Figs. \ref{fig:2} (c) and (d)).

\textcolor{forestgreen}{\emph{\textsf{Conclusion}.}}---
In summary, we propose a general strategy to lift valley degeneracy in MX$_2$ compounds. As exemplified by computational modeling of MoTe$_2$/EuO heterostructure, this approach has several advantages. First, the valley splitting is giant as demonstrated computationally for the MoTe$_2$/EuO heterostructure, and much larger than the band shifts of $\sim0.1$ meV/tesla produced by an external magnetic field.~\cite{Li14c,MacNeill15,Aivazian15,Srivastava15}  Second, with the giant valley splitting, the valley and spin degree of freedom in MX$_2$ can be manipulated statically and dynamically. Third, the induced valley splitting is highly tunable, by controlling the magnetization of the EuO substrate. This is a clearly advantage compared to inducing valley splitting by magnetic doping~\cite{Qi14}. Although we have discussed only the valley splitting for the most stable configuration of the heterostructure, the valley splitting is in fact also prominent for metastable configurations we have examined, as detailed in SI. The giant and tunable valley splitting adds a readily accessible dimension to the valley-spin physics with rich and interesting experimental consequences, and offers a practical avenue for exploring device paradigms based on the intrinsic degrees of freedom of electrons. On a final note, it should be emphasized that the effective Zeeman field experienced by MoTe$_2$ is substantially enhanced by the interfacial interaction. The bare magnetic field  on the EuO(111) surface is very small ($\sim$ 0.007 tesla within the MoTe$_2$), which is shall not lead to significant quantum Hall effect with reasonable doping level at typical experimental temperatures.

\textcolor{forestgreen}{\emph{\textsf{Acknowledgements}.}}---
JF would like to thank Xiaoqiang Liu for helping estimate the surface magnetic field. We acknowledges financial support from the National Science Foundation of China (Projects 11204110 and 11174009), China 973 Program (Projects 2013CB921900 and 2012CB921300), PAPD, ALT\&EI,  DOE (DE-FG03-02ER45958, Division of Materials Science and Engineering) and Welch Foundation (F-1255).

\end{document}